\documentclass[12pt,preprint]{aastex}

\usepackage{amssymb}
\usepackage{amsmath}
\usepackage{xspace}
\usepackage{lscape}

\begin{document}

\title{Reactive desorption and radiative association as possible drivers of complex molecule formation in the cold interstellar medium}

\author{A.I. Vasyunin}
\affil{Department of Chemistry, The University of Virginia, Charlottesville, Virginia, USA}\email{anton.vasyunin@gmail.com}
\author{Eric Herbst}
\affil{Departments of Chemistry, Astronomy, and Physics, The University of Virginia, Charlottesville, Virginia, USA}\email{eh2ef@virginia.edu}

\begin{abstract}
The recent discovery of terrestrial-type organic species such as methyl formate and dimethyl ether in the cold interstellar gas has proved that the formation of organic matter in the Galaxy begins at a much earlier stage of star formation than was thought before. This discovery represents  a challenge for astrochemical modelers. The abundances of these molecules cannot be explained by the previously developed ``warm-up'' scenario, in which organic molecules are formed via diffusive chemistry on surfaces of interstellar grains starting at 30~K, and then released to the gas at higher temperatures during later stages of star formation. In this article, we investigate an alternative scenario in which complex organic species are formed via a sequence of gas-phase reactions between precursor species formed on grain surfaces and then ejected into the gas via efficient reactive desorption, a process in which non-thermal desorption occurs as a result of conversion of the exothermicity of chemical reactions into the ejection of products from the surface. The proposed scenario leads to reasonable if somewhat mixed results at temperatures as low as 10 K and may be considered as a step towards the explanation of abundances of terrestrial-like organic species observed during the earliest stages of star formation.
\end{abstract}


\section{Introduction}
Although it has been understood for some time that a combination of ion-molecule and neutral-neutral reactions can explain much of the exotic ``carbon-chain'' chemistry that occurs in the gas phase of cold interstellar cores, both starless and prestellar,  the chemistry of hot cores, in which gaseous terrestrial (partially saturated) organic species of six or more atoms, known as ``complex organic  molecules'', or COMs for short,  are formed at relatively high abundance, is not well understood \citep{HerbstvanDishoeck09}.   Well-known examples of COMs include methanol (CH$_3$OH), dimethyl ether (CH$_3$OCH$_3$), ethyl cyanide (C$_2$H$_5$CN), and methyl formate (HCOOCH$_3$).  Early theories of the formation of these molecules  were based on  a warm gas-phase chemistry during the hot core stage, possibly preceded by a cold ice-mantle chemistry during the prior cold core stage \citep{Brown_ea12,Charnley_ea92,Hasegawa_ea92,Caselli_ea93,Horn_ea04}.  The cold ice chemistry is thought to be dominated by reactions involving weakly-bound atoms such as hydrogen, which can diffuse at low temperatures.  More recently, the chemistry occurring during the warm-up leading to the formation of a hot core has been emphasized.  In this approach,  the chemical synthesis of hot core molecules begins on the ice mantles of grains in cold cores, where species as complex as methanol are synthesized \citep{GarrodHerbst06}.  As the cold core collapses isothermally into a pre-stellar core and then begins to heat up as a protostellar core is formed, the collapsing gas and dust also increase in temperature.  Starting at temperatures of approximately 30 K, the heavy species formed on the 10 K ice begin to diffuse and collide with one another, although they are mainly unreactive.  Photons or energetic particles strike the dust particles and dissociate some of these heavy species into radicals, which are likely to be quite reactive.  For example, methanol can be dissociated  to produce the methoxy radical (CH$_{3}$O) while formaldehyde can form the formyl radical (HCO).  These two radicals can possibly undergo what is known as a recombination reaction to form methyl formate (HCOOCH$_3$).

Many other COMs are likely formed via recombination of radicals on ices during the warm-up period, as discussed by \citet{GarrodHerbst06} and \citet{Garrod_ea08}, who
 showed that large abundances of COMs  in the gas  can be achieved by the stage at which the temperature of the collapsing material has risen to 100 - 300 K, where thermal evaporation takes place.  At temperatures above 30 K but below 100 K, smaller abundances of complex species formed on grains can also be produced via non-thermal desorption as long as the radical-radical recombination reactions can occur.  Indeed, observations in various sources have shown that small abundances of gaseous COMs can be found in cooler sources than hot cores such as the galactic center or along the line of sight to protostars \citep{Oeberg_ea10}.   But, below about 25 K, it would appear that the diffusion of radicals on ice surfaces  can no longer occur sufficiently rapidly to react to form COMs \citep{GarrodHerbst06}.   Thus, based solely on the warm-up radical mechanism, one would not expect to find any molecules of this type in the gas of cold cores such as TMC-1.

Two recent observations of a number of terrestrial-type complex organic molecules in the cold prestellar cores L1689b and B1-b have excited  the astrochemical community \citep{Bacmann_ea12, Cernicharo_ea12} although some more limited work had reached similar conclusions for a variety of sources including TMC-1 regarding methanol and acetaldehyde (CH$_3$CHO).  New observations of other cold cores (Bacmann, private communication(2013)) also show the presence of COMs.  The COMs seen in L1689b and/or B1-b include  methanol (CH$_3$OH), acetaldehyde (CH$_3$CHO), methyl mercaptan (CH$_3$SH), dimethyl ether (CH$_3$OCH$_3$), and  methyl formate (HCOOCH$_3$),  while smaller species include the methoxy radical (CH$_{3}$O), detected in B1-b for the first time, the formyl radical (HCO), formic acid (HCOOH), ketene (CH$_{2}$CO), and formaldehyde (H$_{2}$CO).    Of the complex species, only the production of methanol is well understood; it is formed via non-thermal desorption following  the successive hydrogenation by atomic hydrogen on ice mantles of CO that is accreted from the gas \citep{WatanabeKouchi02}. In this paper, we propose and investigate a scenario for the formation of the complex terrestrial-type organic molecules methyl formate, dimethyl ether, and acetaldehyde in cold cores.   We also report our results for the gas-phase abundances of ketene, formaldehyde, methoxy, and methanol.  The scenario involves an enhanced rate of non-thermal desorption of selected precursor  molecules (e.g. methoxy, formaldehyde, methanol) from cold ice mantles by a process known as reactive desorption, in which the exothermicity of surface chemical reactions is at least partially channeled into kinetic energy needed to break the bond of the product with the ice.  The enhanced desorption rate is followed by gas-phase reactions leading to the COMs, among which are radiative association processes with rate coefficients that are still poorly understood.  Extension of this approach to other COMs is certainly feasible.

The remainder of this paper is organized as follows.  In Section~2 we describe our chemical model, while in Section~3 modeling results are presented and compared with observational data. Section~4 contains a general discussion, while Section~5 is a summary of our work.

\section{Chemical model}
For this study, we utilized a gas-grain chemical model with a network of gas-phase and grain-surface chemical reactions taken from that used by  \citet{VasyuninHerbst13}, which in turn is based on the well-established OSU astrochemical database\footnote{http://www.physics.ohio-state.edu/$^{\sim}$eric/research.html}.  The diffusive surface chemistry is treated by the rate equation technique \citep{Hasegawa_ea92}, in which no distinction is made between the reactive surface of an ice mantle and the inert icy bulk of interstellar grains, which is a noticeable simplification in comparison with the structure of real multilayer interstellar ices \citep{VasyuninHerbst13}.  Since the methoxy radical (CH$_3$O) was detected in the cold core B1-b \citep{Cernicharo_ea12}, we introduced four gas-phase reactions for this species, which is produced during the hydrogenation of CO on grain mantles followed by desorption.  In our network, the methoxy radical and its isomer, the hydroxymethyl radical (CH$_{2}$OH), are not distinguished.  The first two reactions are simple destruction reactions involving the abundant atoms H and O:
\begin{equation}
{\rm CH_{3}O}+{\rm H}\rightarrow{\rm CH_{3}}+{\rm OH},
\label{methoxy-1}
\end{equation}\
and
\begin{equation}
{\rm CH_{3}O}+{\rm O}\rightarrow{\rm H_{2}CO}+{\rm OH}.
\label{methoxy-2}
\end{equation}
Both of these reactions have been studied to some extent in the laboratory for methoxy and hydroxymethyl radicals, as can be seen in the NIST Chemical Kinetics Database\footnote{http://kinetics.nist.gov/kinetics/Search.jsp}.
For the reaction with atomic hydrogen, only the process involving CH$_{2}$OH has a listed and reviewed rate coefficient, which possesses a temperature independent  value of 1.6$\times$10$^{-10}$~cm$^{3}$ s$^{-1}$ in the range 300-2000 K.  We assume that the rate coefficient remains constant at temperatures down to 10 K in our model.  For the reaction with atomic oxygen,  measured and reviewed values both exist at room temperature for CH$_{2}$OH only; we assume that the average result, 1.0$\times$10$^{-10}$~cm$^{3}$ s$^{-1}$, pertains down to 10 K  in our model.

The third and fourth reactions occur between the methoxy and  methyl radicals:
\begin{equation}
{\rm CH_{3}O}+{\rm CH_{3}}\rightarrow{\rm H_{2}CO}+{\rm CH_{4}},
\label{methoxy-3}
\end{equation}
and
\begin{equation}
{\rm CH_{3}O}+{\rm CH_{3}}\rightarrow{\rm CH_{3}OCH_{3}}+h\nu.
\label{methoxy-4}
\end{equation}
The rate coefficient of reaction (\ref{methoxy-3}), also taken from the NIST Chemical Kinetics Database,  has been measured to be  4.0$\times$10$^{-11}$~cm$^{3}$ s$^{-1}$ by several investigators at room temperature and higher.  We assume that this value pertains down to 10 K.   Reaction (\ref{methoxy-4}) is a radiative association, which has not been studied in the laboratory to the best of our knowledge. The system has been studied at high density, where it essentially reaches the collisional limit at room temperature.
The radiative mechanism, which is the low density process relevant to the interstellar medium, is likely to be much slower at room temperature, but to have a dependence on temperature of T$^{-3}$ \citep{Herbst80}  so that by 10 K, reaction (\ref{methoxy-4}) is also quite rapid.  Based on previous analyses, we estimate the rate coefficient for this reaction to be 10$^{-15} (\rm{T/300~K})^{-3}$ cm$^3$ s$^{-1}$ with a large uncertainty. For convenience, information on new and other important reactions is summarized in Table~\ref{tbl:newr}.

The elemental abundances available for gas-grain chemistry in our simulations correspond to the EA1 set from \citet{WakelamHerbst08}. These are so-called ``low metal'' abundances, which are lower by a factor of $\sim$100 for heavy elements in comparison to solar elemental abundances in order to take into account elemental depletion on grains in cold interstellar cores. The ionization rate $\zeta$ (s$^{-1}$) due to the main ionizing agent  --- cosmic rays -- has a standard value of 1.3$\times$10$^{-17}$~s$^{-1}$ for molecular hydrogen.  The initial abundances are all atomic except for hydrogen, which starts in the form of H$_{2}$.  The gas-to-dust mass ratio is 0.01.

To treat the diffusive surface chemistry, we assume that the ice mantle surrounding interstellar grains can be represented as a two-dimensional lattice with a periodic potential. Wells of the potential are the binding sites for atomic and molecular adsorbates. Due to thermal hopping, species can move across the grain surface and react with each other. In the case of atomic and molecular hydrogen, quantum tunneling through the potential barriers is also considered as a source of surface mobility. The width of a potential barrier between two binding sites is assumed to be 1~\AA. The height $E_{\rm b}$ of a barrier against diffusion for a species  is assumed to be \(1/2\) of the desorption energy $E_{\rm D}$. In our simulations, we consider spherical grains of a single size of 10$^{-5}$~cm with $\sim$10$^{6}$ binding sites. Note that our network is inherited from previous studies in which complex organic molecules were formed at higher temperatures via diffusive surface chemistry. It contains surface radical-radical chemical reactions that lead to the formation of COMs. Since at 10~K, the mobility of molecular species on surfaces is negligible, these reactions do not contribute to the formation of COMs in this study. The only efficient diffusive reactants on grain surfaces at 10~K are atoms such as hydrogen, oxygen and nitrogen, especially atomic hydrogen, which can be an efficient tunneler. For example, hydrogen atoms that accrete onto dust particle surfaces are able to hydrogenate CO ice all the way to methanol ice  at 10~K \citep{Charnley_ea97,WatanabeKouchi02}:
\begin{equation}
{\rm CO}\rightarrow{\rm HCO}\rightarrow{\rm H_{2}CO}\rightarrow{\rm H_{3}CO}\rightarrow{\rm CH_{3}OH}.
\label{CH3OHformation}
\end{equation}

The chemistry in the gas phase is connected to grain-surface chemistry via accretion and desorption. The accretion probability for all neutral species is set to unity except for helium, for which it is taken to be zero because it is chemically inert in surface chemistry and has a very low desorption energy, so that it would thermally desorb rapidly anyway. Molecular ions are not allowed to stick on grains. There are four types of desorption included in our model: thermal evaporation \citep{Hasegawa_ea92}, photodesorption \citep{Fayolle_ea11}, desorption due to grain heating by cosmic ray bombardment \citep{HasegawaHerbst93}, and reactive desorption \citep{Garrod_ea07}. In cold dark cloud cores, which are well shielded from UV radiation and have very low temperatures of $\approx$10~K, thermal evaporation is nil except for He, H, and H$_{2}$ while photodesorption can only occur slowly via photons produced by energetic electrons.  Desorption due to cosmic rays has limited efficiency and has only been proven to be important on large timescales for species such as CO \citep{HasegawaHerbst93}. Constraints on the efficiency of reactive desorption, in which the products of an exothermic surface reaction are immediately ejected into the gas, are poor. To the best of our knowledge, there are no experimental constraints on its efficiency except for the case of H$_{2}$ \citep{Katz_ea99}, and only a simple statistical theory is available for larger products \citep{Garrod_ea07}. In this study, we consider three different efficiencies for reactive desorption: 0\%, 1\%, and 10\% per exothermic reaction on a grain surface.  The 1\% efficiency is sufficient to  explain observed fractional abundances of methanol ($\approx 10^{-9}$) in cold cores \citep{Garrod_ea07}.  The 10\% value is equal to the highest efficiency considered by  \citet{Garrod_ea07}.  Our models with these three different efficiencies for reactive desorption are labeled M0, M1, and M10.

All chemical models are run for 10$^{6}$ yr, which is slightly longer than so-called ``early time'', at which the ion-molecule chemistry best reproduces the carbon-chain species observed in cold cores.  We utilize physical conditions typical for sources such as L1689b and B1-b, where COMs were found at low temperature: T=10~K,  a proton density $n_{\rm H}$=10$^{5}$~cm$^{-3}$, and a visual extinction $A_{\rm V}$=10. In our analysis, we focus on the time span 10$^{5}$---10$^{6}$ yr.

\section{Results}
\begin{figure}
  \includegraphics[height=0.60\textwidth, angle=90]{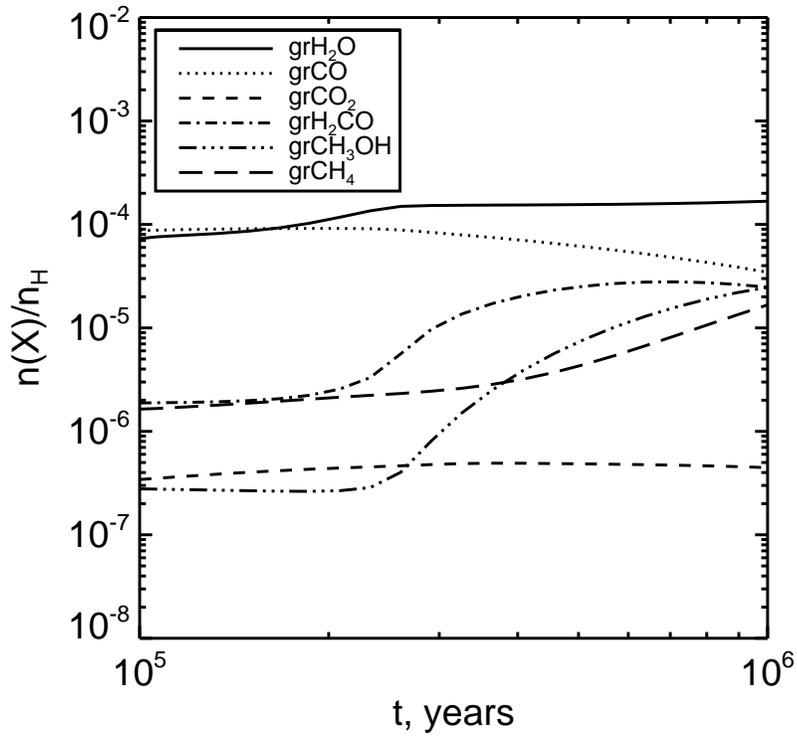}\vspace{5mm}
  \caption{Abundances vs. time of water and major carbon-bearing ice compounds in the model M10.  The prefix ``gr'' denotes species on and in ice mantles of interstellar grains.}
  \label{fgr:ice}
\end{figure}

Because the calculated ice mantle composition in all three models is quite similar, we present the composition  for only one model -- M10 -- in Figure \ref{fgr:ice}. The similarity of the ice composition in all considered models arises because in this work reactive desorption affects the rates of all surface reactions by a relatively small factor of at most 10\%. As such, it does not change considerably the balance between different surface chemical processes, and the resulting ice composition remains almost unchanged. As expected, our chemical model produces a typical ice composition which is similar to that calculated in a number of previous studies
\citep{GarrodPauly11,VasyuninHerbst13}, and which is also generally similar to observed ice abundances towards protostars \citep{Oeberg_ea11}. The major ice component is water,  which has a nearly constant fractional abundance of $\sim$10$^{-4}$ after 10$^{5}$ yr. The abundances of  the carbon-bearing major ice components change with time more significantly. Carbon monoxide  accretes from the gas phase  and at 10$^{5}$ yr possesses an abundance on grain surfaces similar to that of water.   By 10$^{6}$ yr,  its abundance has decreased by approximately a factor of 3 as it is gradually converted to methanol. Correspondingly, the abundances of  formaldehyde, methanol and methane rise with time by an order of  magnitude or more to reach values of roughly ten to thirty percent of the abundance of water ice. Another major ice constituent observed in ices towards star-forming regions, carbon dioxide, has a very small fractional abundance of $\sim 4\times10^{-7}$ in our model, despite a much larger observed abundance. This discrepancy arises because carbon dioxide is probably formed during other stages of star formation when dust grains have higher temperatures of $\sim$20~K \citep[e.g.,][]{GarrodPauly11, VasyuninHerbst13}.

In Figure \ref{fgr:gas}, abundances of selected gas-phase species are presented for models M0, M1, and M10, with formaldehyde, methanol, and methoxy in the left panels and methyl formate, ketene (CH$_2$CO), dimethyl ether, and acetaldehyde in the right panels. The upper row of panels represents results for M0 (no reactive desorption), the middle row results for M1 (1\% reactive desorption efficiency), and the lower row results for model M10 (10\% reactive desorption efficiency). The results for M1 are similar to those of the best-fit model in the earlier paper of \citet{Garrod_ea07}.   Fractional abundances of all these species observed in L1689b and B1-b are shown in the two rightmost columns of Table~\ref{tbl:abu}. We derived the observed fractional abundances in these sources by dividing the observed column densities as presented in \citet{Bacmann_ea12} and  \citet{Cernicharo_ea12}  by the column densities of molecular hydrogen in L1689b and B1-b.   The observed fractional abundances in L1689b range from approximately 10$^{-10}$ to
10$^{-9}$ whereas those in B1-b range from approximately 10$^{-11}$ to more than 10$^{-9}$.  Given the uncertainties both in observed abundances and in our astrochemical models, we consider a model to explain successfully the observed abundance of a species if the modeled abundance differs from the observed abundance by not more than one order of magnitude \citep{Vasyunin_ea04, Vasyunin_ea08, Wakelam_ea10}.

\subsection{M0: a model without reactive desorption}
Formaldehyde is mainly produced by the gas-phase reaction ${\rm O}+{\rm CH_{3}}\rightarrow{\rm H_{2}CO}+{\rm H}$, while ketene (${\rm CH_{2}CO}$) has two paths of formation. First, it is produced in the dissociative recombination of ${\rm C_2H_3O^+}$, which in turn is created in the radiative association reaction between the second most abundant gas-phase molecule ${\rm CO}$ and the simple ion ${\rm CH_3^+}$:
\begin{equation}
{\rm CO}+{\rm CH_{3}^{+}}\rightarrow{\rm C_{2}H_{3}O^{+}} + h\nu.
\label{ketene-1}
\end{equation}
Secondly, ketene is produced in the neutral-neutral reaction
\begin{equation}
{\rm O}+{\rm C_{2}H_{3}}\rightarrow{\rm CH_{2}CO}+{\rm H}.
\label{ketene-2}
\end{equation}
Other observed organic molecules in the model M0 have very low abundances, because they do not have gas-phase routes of formation efficient  enough at 10~K to compete with accretion.

\subsection{M1 and M10: models with reactive desorption}
\begin{figure}
  \includegraphics[height=0.75\textwidth, angle=90]{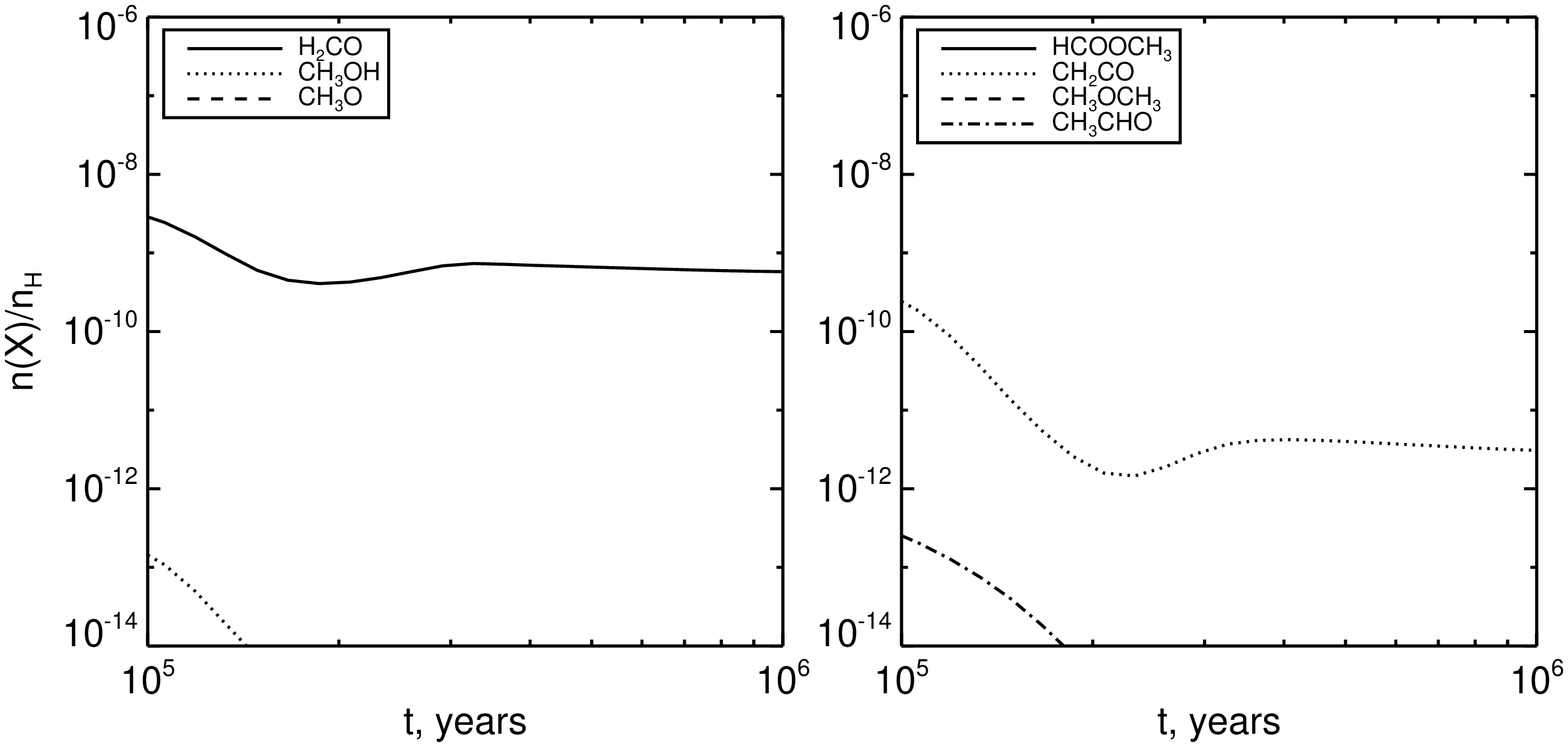}
  \includegraphics[height=0.75\textwidth, angle=90]{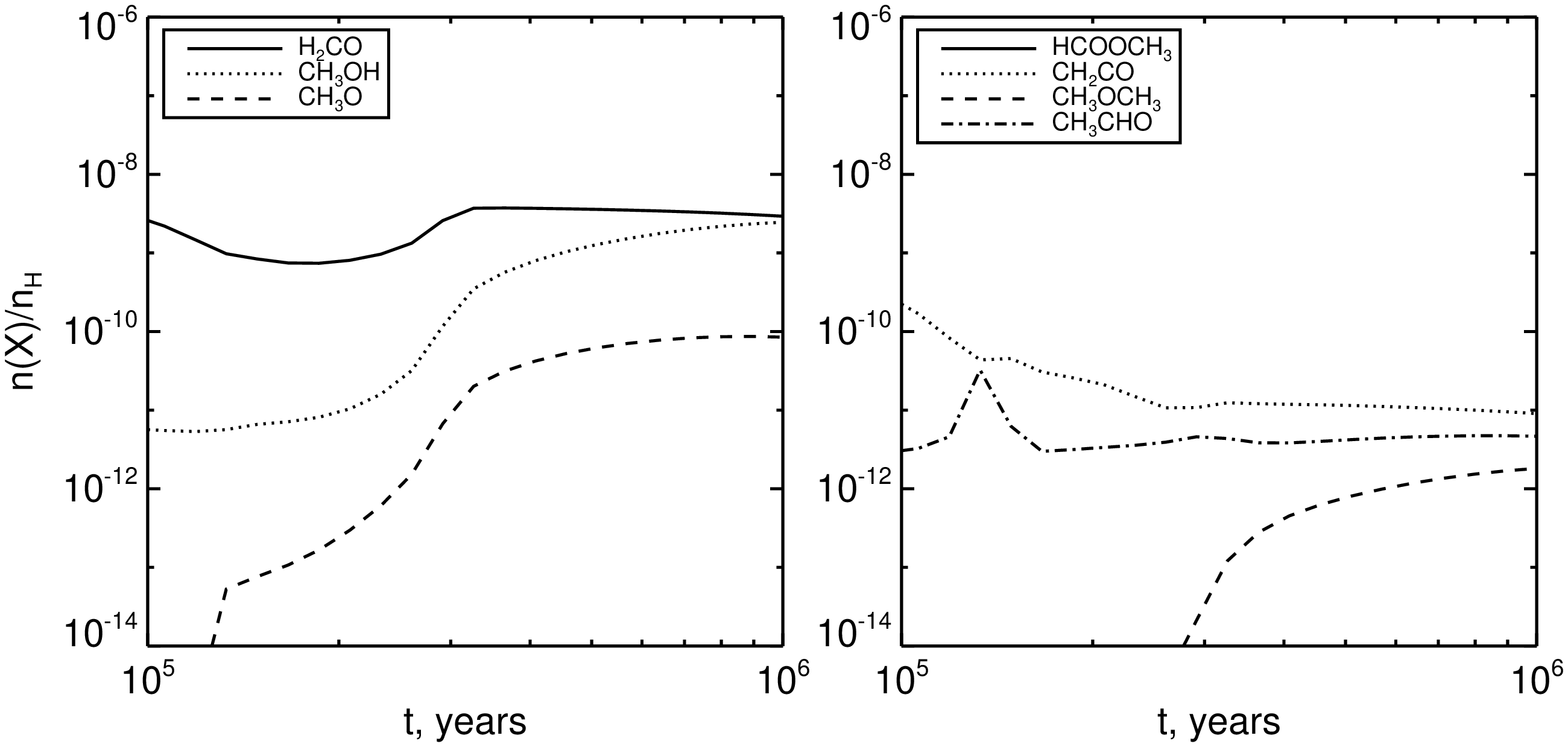}
  \includegraphics[height=0.75\textwidth, angle=90]{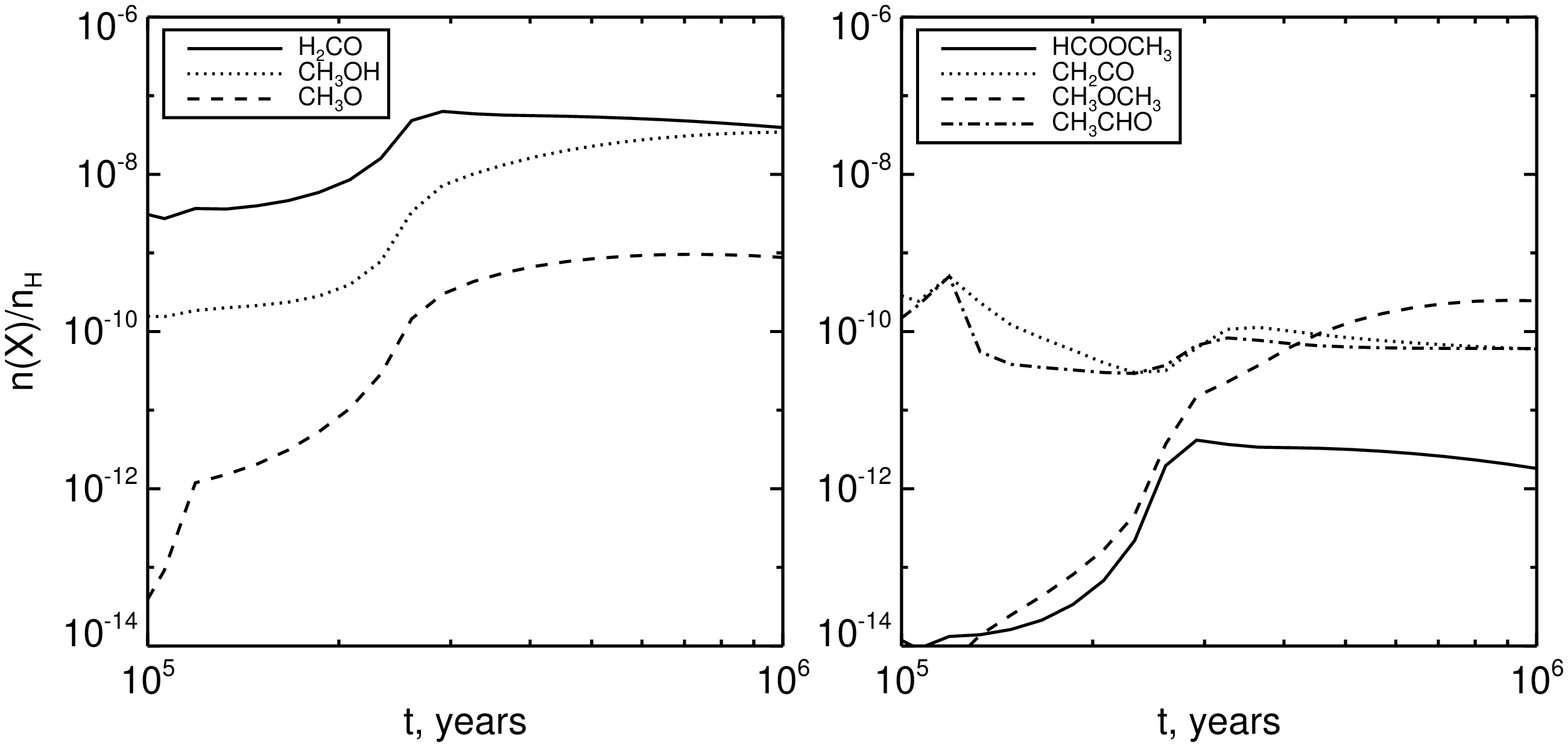}\vspace{5mm}
  \caption{Gas-phase abundances vs time for simple and complex organic species  in models M0 (top row), M1 (middle row), and M10 (bottom row). }
  \label{fgr:gas}
\end{figure}

Our results change drastically when reactive desorption is enabled.  For model M1, as can be seen in the left middle row of Figure \ref{fgr:gas},   formaldehyde and methanol reach peak abundances in excess of 10$^{-9}$, while the methoxy radical reaches a peak abundance of 10$^{-10}$.
As discussed earlier, methanol is not formed in the gas, but is formed by surface hydrogenation of CO followed by desorption
\citep{Garrod_ea07}. This is not only true for ${\rm CH_{3}OH}$ but also for intermediate products of the reaction chain in eq.~(\ref{CH3OHformation}), and other products of surface reactions. As such, reactive desorption becomes a major source of methanol and the methoxy radical in the gas phase. The abundance of formaldehyde is affected to a lesser extent because it is already produced efficiently by gas-phase processes. The increase in the gas-phase abundances of methoxy, methanol and formaldehyde after 3$\times$10$^{5}$ yr is caused by the increased abundances of these species in the ice mantle, as can be seen in Figure \ref{fgr:ice}.

For the molecules in the right middle panel, model M1 also shows enhanced abundances over M0. While the final abundance of ketene (${\rm CH_{2}CO}$)  increases only by a factor of three, the abundance of acetaldehyde (${\rm CH_{3}CHO}$) is increased by four orders of magnitude, and the abundance of dimethyl ether (${\rm CH_{3}OCH_{3}}$) is increased by more than ten orders of magnitude. The fractional abundances of both species with respect to hydrogen now exceed 10$^{-12}$ towards 10$^{6}$ yr.
The additional acetaldehyde is produced via the gas-phase reaction $\rm{O} + \rm{C_{2}H_{5}} \rightarrow \rm{CH_{3}CHO} + \rm{H}$. The  \rm{C$_2$H$_5$} product is ejected into the gas-phase via reactive desorption as an intermediate product of the chain of surface reactions that hydrogenate  $\rm{C_2}$ into ${\rm C_2H_6}$:
\begin{equation}
\rm{C_2}\rightarrow\rm{C_{2}H}\rightarrow\rm{C_{2}H_{2}}\rightarrow\rm{C_{2}H_{3}}\rightarrow\rm{C_{2}H_{4}}\rightarrow\rm{C_{2}H_{5}}\rightarrow\rm{C_{2}H_{6}}.
\label{C2H5form}
\end{equation}
Gas-phase dimethyl ether, on the other hand,  is mainly produced via the gas-phase radiative association reaction shown in eq.~(\ref{methoxy-4}).
Although inclusion of the ``standard'' reactive desorption with an efficiency of 1\% in the model allows us to produce complex organic molecules in much higher abundances than with no reactive desorption at all, some of the predicted abundances are still quite low. For example, methyl formate (${\rm HCOOCH_{3}}$) has an abundance in model M1 below 10$^{-14}$  in the entire time span 10$^{5}$--10$^{6}$ yr, which is  3--4 orders of magnitude below the observed values.

The results produced by the model M10, with an efficiency for reactive desorption of 10\%,  are presented in the bottom row of Figure~\ref{fgr:gas}. In this model, the abundances of all depicted species are enhanced in comparison with model M1 and especially with model M0. In model M10, the contribution to gas-phase H$_2$CO from the chain of surface reactions in reaction~(\ref{CH3OHformation}) via reactive desorption is an order of magnitude larger than the gas-phase formation routes, which leads to an increase in the peak fractional abundance of ${\rm H_{2}CO}$ to 6$\times$10$^{-8}$. The abundances of methoxy and methanol are increased by  an order of magnitude in comparison with model M1, and reach 8.7$\times$10$^{-10}$ and 3.4$\times$10$^{-8}$, respectively, at their maximum.

As can be seen on the right panel in the bottom row of Figure~\ref{fgr:gas}, the abundances of ketene and complex organic species are increased in model M10 in comparison with model M1 by different factors. The smallest enhancement in abundance, a factor of 7 at 10$^{6}$ yr, is exhibited  by ketene. The order of magnitude increase in the abundance of gas-phase methanol throughout the time range shown leads to an increase in the abundance of ${\rm CH_{3}^{+}}$  via the reaction ${\rm H_{3}^{+}}+{\rm CH_{3}OH}\rightarrow{\rm CH_{3}^{+}}+{\rm H_{2}O}+{\rm H_{2}}$.
The methyl ion can then react with CO in  reaction~(\ref{ketene-1}) to produce ${\rm C_{2}H_{3}O^{+}}$, and subsequently ${\rm CH_{2}CO}$.   An order-of-magnitude increase in the abundance of acetaldehyde is caused by the increased abundance of ${\rm C_{2}H_{5}}$, which is ejected into the gas via efficient reactive desorption from the surface reaction chain in reaction~(\ref{C2H5form}).

The abundance of dimethyl ether in model M10 is increased by two orders of magnitude in comparison with model M1 because the abundances of both gaseous reactants (${\rm CH_{3}}$ and ${\rm CH_{3}O}$) in reaction~(\ref{methoxy-4}) are increased in model M10 by an order of magnitude. The increase in gaseous ${\rm CH_{3}O}$ derives from the increase in efficiency of reactive desorption in model M10. In turn, ${\rm CH_{3}}$ in model M10 is mainly produced in the gas-phase via reaction~(\ref{methoxy-1}). Another route for the formation of ${\rm CH_{3}OCH_{3}}$ is via the dissociative recombination of protonated dimethyl ether, which is produced in the reaction of methanol with protonated methanol:
\begin{equation}
{\rm CH_{3}OH_{2}^{+} + CH_{3}OH  \rightarrow CH_{3}OHCH_{3}^{+} + H_{2}O  }
\label{CH3OCH3-second}
\end{equation}

The strongest abundance enhancement in switching from model M1 to model M10  occurs for methyl formate, which increases from 4$\times$10$^{-15}$ to 1.8$\times$10$^{-12}$ at 10$^{6}$ yr. The most efficient gas-phase formation route for ${\rm HCOOCH_{3}}$ in our model is via dissociative recombination of ${\rm H_{2}COHOCH_{2}^{+}}$. The principal formation route of this species occurs via the radiative association reaction \citep{Horn_ea04}
\begin{equation}
{\rm H_{2}COH^{+} +  H_{2}CO \rightarrow H_{2}COHOCH_{2}^{+}+h\nu .}
\label{HCOOCH3-form}
\end{equation}
The product is not normal protonated methyl formate, which has the structure ${\rm HC(OH)OCH_{3}^{+}}$. Unfortunately there is a transition state barrier between the product of reaction~(\ref{HCOOCH3-form}) and protonated methyl formate \citep{Horn_ea04}. We assume, as did \citet{Horn_ea04}, that dissociative recombination of the product of reaction~(\ref{HCOOCH3-form}) can also form methyl formate + H despite the change in structure required. Reaction~(\ref{HCOOCH3-form}) is similar to reaction~(\ref{CH3OCH3-second}) in the sense that its rate depends quadratically on the abundance of gas-phase formaldehyde.  Consequently, the gas-phase abundance of ${\rm HCOOCH_{3}}$ in the model M10  becomes comparable although still somewhat lower than the observed values at times exceeding $3 \times 10^{5}$ yr.  If the dissociative recombination reaction does not lead to methyl formate, another process that can produce protonated methyl formate is the radiative association reaction \citep{Horn_ea04}
\begin{equation}
{\rm CH_{3}^{+}  +  HCOOH  \rightarrow  HC(OH)OCH_{3}^{+}  +   h\nu,}
\end{equation}
followed by dissociative recombination.  This reaction, however, is not as efficient in producing methyl formate in our current models.

The general behavior of models with reactive desorption is similar to that described in \cite{Garrod_ea07}. These authors found that at  $\le$2$\cdot$10$^{6}$ yr, the abundances of the majority of gas-phase species are not affected significantly by the inclusion of reactive desorption. The exceptions are several hydrogenated species such as methanol and formaldehyde. In this study, we observe a similar picture extended by the enhanced abundances of COMs, descendants of H$_{2}$CO and CH$_{3}$OH, which were not considered in \cite{Garrod_ea07}.

\subsection{Comparison with observations}
It is interesting to investigate how well our models reproduce the observed abundances of complex organic species and their precursors. To quantify the level of agreement, we calculated individual confidence parameters $\kappa_i$ for each $i$-th species versus time in models M1 and M10 according to the prescription in \citet{Garrod_ea07} and then took the mean of the individual confidence parameters as the criterion of overall agreement. The formula for each individual confidence parameter is
\begin{equation}
\kappa_i=erfc\left(\frac{|log(X_i)-log(X_i^{obs})|}{\sqrt{2}\sigma} \right),
\end{equation}
where  $erfc$ is the complementary error function, $X_i$ and $X_i^{obs}$ are the modeled and observed abundances of species $i$ at time $t$, and the standard deviation $\sigma$ is assumed to equal unity, which means that we estimate the uncertainty in observed abundances to be one order of magnitude. With this definition, $\kappa_i$  is equal to unity when the observed and model abundances  are equal, and quickly goes towards a limit of zero with increasing discrepancy between these two values. For example, when the difference is one order of magnitude, $\kappa_i$ = 0.32, and when the difference is two orders of magnitude, $\kappa_i$ = 0.046. Model M0 is excluded from this comparison as its results are obviously too far from observational values.

\begin{figure}
\includegraphics[height=0.70\textwidth, angle=0]{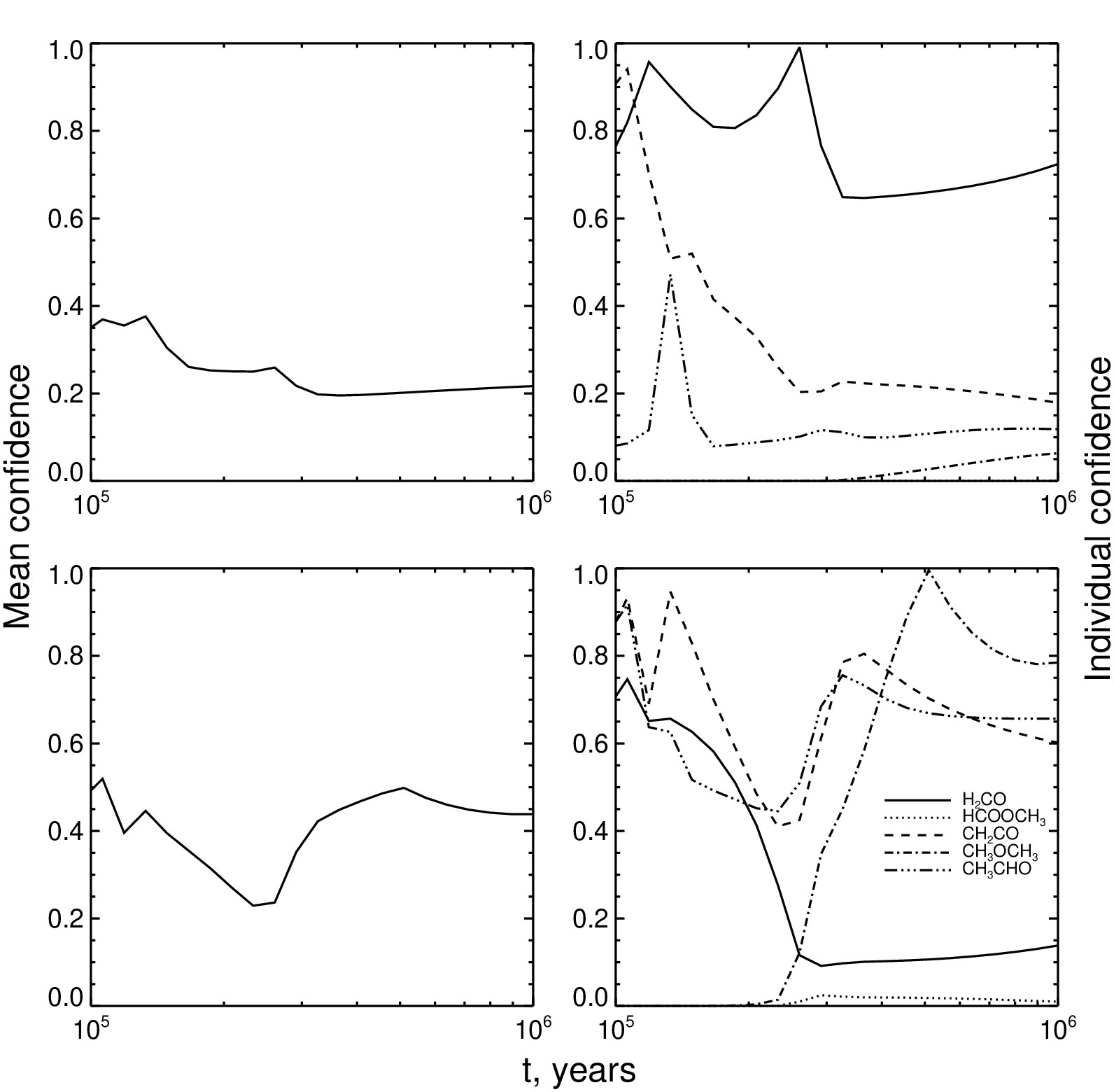}
\caption{Mean and individual  confidence levels for the species in models M1 (upper panels) and M10 (lower panels) in the case of L1689b. The higher the mean confidence level, the better the agreement between modeled and observed abundances.}\label{conf-l1689b}
\end{figure}

\begin{figure}
\includegraphics[height=0.70\textwidth, angle=0]{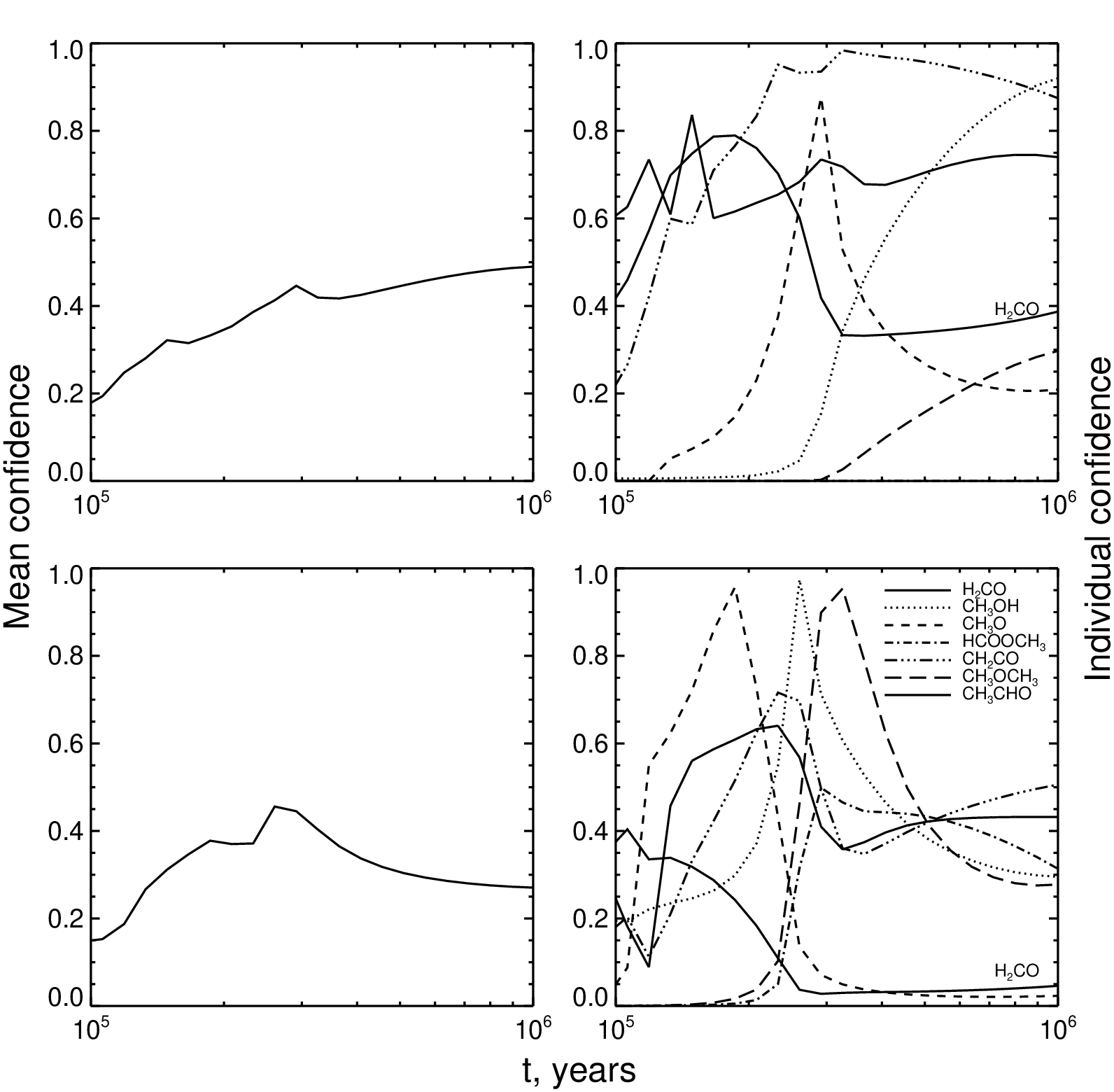}
\caption{Mean and individual  confidence levels for the species in models M1 (upper panels) and M10 (lower panels) in the case of B1-b. The higher the mean confidence level, the better the  agreement between modeled and observed abundances.}\label{conf-b1-b}
\end{figure}

The molecules shown in Table~\ref{tbl:abu} are used in the analysis. In addition to these molecules, several other simple species were observed towards B1-b and L1689b such as CO, HCO$^{+}$, N$_{2}$H$^{+}$, NH$_{3}$, CS and HNCO/HCNO \citep[][]{Hirano_ea99, Bacmann_ea02, Marcelino_ea09, Oeberg_ea10}. We did not include them in the comparison set for two reasons. First, the abundances of these species come from different sets of observations, and their inclusion will make the set heterogeneous. Secondly, the abundances of these species are only slightly different in our models with and without reactive desorption. The resulting individual and  mean confidence parameters for L1689b and B1-b are shown in Figures~\ref{conf-l1689b} and \ref{conf-b1-b}, respectively.  For L1689b, model M1 reaches its best agreement at 1.3$\times$10$^{5}$ yr and model M10 at 5.1$\times$10$^{5}$ yr (if we exclude for M10 a small cusp at much earlier time that has a similar extent of agreement). In the case of B1-b, the time of best agreement time for model M1 is $1.0 \times10^{6}$ yr, while for model M10 it is 2.6$\times$10$^{5}$ yr.  At these times, the average agreement is better than one order of magnitude.   The individual molecular abundances at these best times are shown in Table~\ref{tbl:abu}. Because there are two more organic species detected in B1-b than in L1689b, we choose B1-b for detailed discussion.

At the time of best agreement for M1, the modeled abundances for the two precursor species  ${\rm CH_{2}CO}$ and  ${\rm H_{2}CO}$, and the two complex species ${\rm CH_{3}OH}$  and ${\rm CH_{3}CHO}$, out of seven species,  differ from the observational values by less than one order of magnitude. For one species, ${\rm CH_{3}OCH_{3}}$, the difference between modeled and observed abundances just slightly exceeds one order of magnitude.  At the same time, model M1 somewhat overproduces methoxy and significantly underproduces methyl formate. Model M10 exhibits its  best agreement with observations of B1-b significantly earlier.  While it manages to reproduce the abundances of all complex organic species including ${\rm HCOOCH_{3}}$,  the modeled abundance of ${\rm H_{2}CO}$ is  two orders of magnitude higher than the observed value. The abundance of ${\rm CH_{3}O}$ is also significantly higher than the observed value.

A possible reason for the overly high calculated abundance of methoxy in both the M1 and M10 models  has been given by \citet{Cernicharo_ea12}, who showed that that on grains  the isomer ${\rm CH_{2}OH}$  is produced more efficiently than ${\rm CH_{3}O}$ and released into the gas phase. This isomer has different spectral properties, and was not observed in the gas phase. The authors proposed an alternative formation route for methoxy via the gas-phase reaction
\begin{equation}
\rm {CH_{3}OH+OH  \rightarrow  CH_{3}O +  H_{2}O}.
\label{CH3Oprod}
\end{equation}
Although our chemical model currently does not discriminate between ${\rm CH_{3}O}$ and ${\rm CH_{2}OH}$,  we estimate that if methoxy is only produced by reaction~(\ref{CH3Oprod}) at its newly determined rate coefficient, then the abundance of methoxy will be in much closer agreement with observations, at least in case of B1-b.  But, one must remember that since methoxy is a precursor species to complex organic molecules, lessening its abundance will also lessen the abundance of the complex molecules formed from it, especially dimethyl ether.

\begin{figure}
  \includegraphics[height=0.90\textwidth, angle=90]{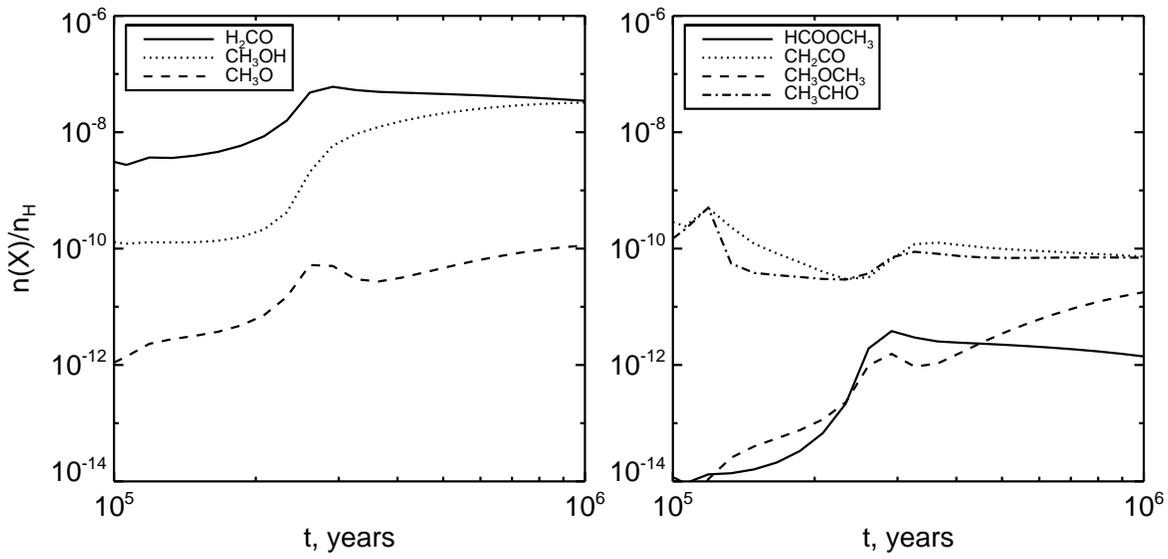}\vspace{5mm}
  \caption{Abundances vs time of complex organic species and their precursors in the altered model M10,  with reactive desorption of methoxy radicals switched off and reaction (\ref{CH3Oprod}) enabled.}
  \label{fgr:gas-no-ch3o-rdes}
\end{figure}

We show the organic abundances vs time for an altered model  M10, in which  the gas-phase route to methoxy is included and  the reactive desorption route is removed,  in Figure~\ref{fgr:gas-no-ch3o-rdes}.  The results can be compared with the abundances obtained with the original model M10, which are shown in the bottom panel of Figure \ref{fgr:gas}.
The time of the best agreement between observed values for B1-b and results of the altered model M10 is the same as for the original model, but at later times, the level of agreement remains at almost the same value until 10$^{6}$ yr.
For B1-b, the new fractional abundance of $\rm{CH_{3}O}$ at the time of best agreement is 5$\times$10$^{-11}$,  which is closer to the observed value, although still too large by an order of magnitude. This fact may imply that the rate coefficient of  reaction (\ref{CH3Oprod}) is somewhat overestimated, which is supported by older estimations. With the high value for the rate coefficient of reaction~(\ref{CH3Oprod}), the new abundance of ${\rm CH_{3}OCH_{3}}$ at the moment of best agreement for B1-b is 1 $\times$ 10$^{-12}$, but it increases steadily, reaching a value of 2$\times$10$^{-11}$  at 10$^{6}$ yr.

Finally, it is interesting to note that in B1-b the observed abundances of all organic species we consider are lower than in L1689b, including formaldehyde. If future observations  confirm that this correlation is not caused by systematic errors between observational data sets, the lower abundances in B1-b argue in
 favor of our scenario for the formation of complex molecules in cold clouds, because our scenario implies a strong correlation between the abundances of ${\rm H_{2}CO}$ and COMs.

\section{Discussion}\label{discussion}
We have considered one approach to understanding how terrestrial-type organic molecules can be synthesized in cold starless and prestellar cores. In our approach, reactive desorption succeeds in removing precursor molecules efficiently from the ice mantles  of cold dust grains into the gas, where subsequent reactions produce the organic molecules in reasonable, if not perfect, agreement with the small observed abundances detected in the cold  prestellar cores L1689b and B1-b.  Radiative association reactions in the gas play an important role in the synthesis of the complex organic molecules methyl formate and dimethyl ether. Both reactive desorption and radiative association reactions are still poorly understood processes in general.   Moreover, the product for the proposed radiative association reaction that leads to the formation of protonated methyl formate most likely yields a different structure than desired, which may or may not form methyl formate in its lowest conformer when the ion undergoes a dissociative recombination with electrons.

Although we have discussed our model results mainly for observed COMs, our model can also be tested by its predictions for the abundances of other complex species, such as ethyl cyanide (C$_2$H$_5$CN), which is a well-known weed in hot cores.   Under cold core conditions, our model first produces ethyl cyanide in the ice mantle by hydrogenation of the CCCN radical \citep{Hasegawa_ea92, Caselli_ea93}.   Reactive desorption then results in a considerable abundance of this species in the gas phase at selected times.   For example, in Model M1, the fractional abundance of ethyl cyanide peaks at over 10$^{-10}$ at a time near $2 \times 10^{5}$ yr, while for Model M10, the peak value approaches a fractional abundance of 10$^{-8}$ at the same time. At the times of best agreement for models M1 and M10 with B1-b,  the calculated fractional abundances of ethyl cyanide are somewhat under 10$^{-12}$ and somewhat over 10$^{-10}$ respectively, while for L1689b, these numbers are somewhat over 10$^{-10}$ and near 10$^{-11}$ .

There are perhaps other  complementary approaches to the production of COMs in cold cores, where the dust surfaces are typically too cold for the surface radical-radical photochemistry suggested by \citet{GarrodHerbst06}.   One possibility is that we are missing a large number of thermal surface/ice reactions that can occur at 10 K, typically by the diffusion of atoms somewhat heavier than hydrogen.  A figure showing the many species that might be synthesized in this way was produced by S. Charnley (2009, private communication)  and can  be found in \citet{HerbstvanDishoeck09}.   Another possibility is that the dust particles are not always at 10 K, and that a variety of effects cause a temporary increase in temperature high enough to allow surface radicals to diffuse and react with one another.  It is even possible that photodissociation or particle bombardment processes can cause temporary non-thermal motions of species so as to promote the rates of non-thermal diffusive reactions \citep{Occhiogrosso_ea11, KimKaiser12}.  Finally, tunneling from one lattice site to another for species other than atomic hydrogen will promote more rapid surface/ice reactions at lower temperatures than if the only diffusive mechanism for species heaver than H is simple thermal hopping.  We hope to investigate some of these possibilities in the near future.

\section{Summary}\label{summary}
In this study, we have proposed and examined a scenario of formation of complex organic molecules recently found in the cold cores L1689b and B1-b with gas temperatures of 10~K. The proposed scenario is based on the assumption that complex organic species in the cold regions are formed via gas-phase ion-molecular and neutral-neutral chemistry from simpler precursors such as formaldehyde and methanol. These precursors are formed on icy surfaces of interstellar grains, and then ejected into the gas via reactive desorption. Because of our lack of knowledge of the efficiency of reactive desorption, we considered three models:   M0, M1, and M10, in which the efficiency of reactive desorption per exothermic surface reaction is 0\%, 1\%, and 10\%, respectively.

We found that the proposed scenario gives somewhat mixed agreement with observations of COMs and their precursors. Both models with non-zero reactive desorption exhibit much better agreement with observations than the model without reactive desorption. The best agreement for the whole set of species is exhibited by model M10, which possesses the highest efficiency of reactive desorption. On the other hand, this model tends to overestimate gas-phase abundances of formaldehyde in comparison to observations by $\sim$ two orders of magnitude. Such a discrepancy can be partly but not fully explained by uncertainties in our chemical model, particularly in the rate coefficients of chemical reactions.  This fact may imply that the proposed scenario of formation of complex organic species via gas-phase reactions driven by reactive desorption is only partially responsible for the  formation of these molecules, and should be assisted by other routes such as processes that warm up grains temporarily or that permit non-thermal ice reactive processes.

\acknowledgements
The authors wish to thank the anonymous referee for valuable comments, which helped us to improve the manuscript.  E. H. acknowledges the support of the National Science Foundation for his astrochemistry program, and  support from the NASA Exobiology and Evolutionary Biology program through a subcontract from Rensselaer Polytechnic Institute.  This research has made use of NASA's Astrophysics Data System.


\begin{thebibliography}{36}
\expandafter\ifx\csname natexlab\endcsname\relax\def\natexlab#1{#1}\fi

\bibitem[{Anicich {et~al.}(2003)Anicich, Aeronautics, Administration, \&
  (U.S.)}]{Anicich03}
Anicich, V., Aeronautics, U. S.~N., Administration, S., \& (U.S.), J. P.~L.
  2003, JPL Publication, Vol. 03-19, An Index of the Literature for Bimolecular
  Gas Phase Cation-Molecule Reaction Kinetics (Jet Propulsion Laboratory,
  National Aeronautics and Space Administration, Pasadena, CA)

\bibitem[{{Bacmann} {et~al.}(2002){Bacmann}, {Lefloch}, {Ceccarelli},
  {Castets}, {Steinacker}, \& {Loinard}}]{Bacmann_ea02}
{Bacmann}, A., {Lefloch}, B., {Ceccarelli}, C., {Castets}, A., {Steinacker},
  J., \& {Loinard}, L. 2002, \aap, 389, L6

\bibitem[{{Bacmann} {et~al.}(2012){Bacmann}, {Taquet}, {Faure}, {Kahane}, \&
  {Ceccarelli}}]{Bacmann_ea12}
{Bacmann}, A., {Taquet}, V., {Faure}, A., {Kahane}, C., \& {Ceccarelli}, C.
  2012, \aap, 541, L12

\bibitem[{{Brown} {et~al.}(1988){Brown}, {Charnley}, \& {Millar}}]{Brown_ea12}
{Brown}, P.~D., {Charnley}, S.~B., \& {Millar}, T.~J. 1988, \mnras, 231, 409

\bibitem[{{Caselli} {et~al.}(1993){Caselli}, {Hasegawa}, \&
  {Herbst}}]{Caselli_ea93}
{Caselli}, P., {Hasegawa}, T.~I., \& {Herbst}, E. 1993, \apj, 408, 548

\bibitem[{{Cernicharo} {et~al.}(2012){Cernicharo}, {Marcelino}, {Roueff},
  {Gerin}, {Jim{\'e}nez-Escobar}, \& {Mu{\~n}oz Caro}}]{Cernicharo_ea12}
{Cernicharo}, J., {Marcelino}, N., {Roueff}, E., {Gerin}, M.,
  {Jim{\'e}nez-Escobar}, A., \& {Mu{\~n}oz Caro}, G.~M. 2012, \apjl, 759, L43

\bibitem[{{Charnley} {et~al.}(1992){Charnley}, {Tielens}, \&
  {Millar}}]{Charnley_ea92}
{Charnley}, S.~B., {Tielens}, A.~G.~G.~M., \& {Millar}, T.~J. 1992, \apjl, 399,
  L71

\bibitem[{{Charnley} {et~al.}(1997){Charnley}, {Tielens}, \&
  {Rodgers}}]{Charnley_ea97}
{Charnley}, S.~B., {Tielens}, A.~G.~G.~M., \& {Rodgers}, S.~D. 1997, \apjl,
  482, L203

\bibitem[{{Fayolle} {et~al.}(2011){Fayolle}, {Bertin}, {Romanzin}, {Michaut},
  {{\"O}berg}, {Linnartz}, \& {Fillion}}]{Fayolle_ea11}
{Fayolle}, E.~C., {Bertin}, M., {Romanzin}, C., {Michaut}, X., {{\"O}berg},
  K.~I., {Linnartz}, H., \& {Fillion}, J.-H. 2011, \apjl, 739, L36

\bibitem[{{Garrod} \& {Herbst}(2006)}]{GarrodHerbst06}
{Garrod}, R.~T. \& {Herbst}, E. 2006, \aap, 457, 927

\bibitem[{{Garrod} \& {Pauly}(2011)}]{GarrodPauly11}
{Garrod}, R.~T. \& {Pauly}, T. 2011, \apj, 735, 15

\bibitem[{{Garrod} {et~al.}(2007){Garrod}, {Wakelam}, \&
  {Herbst}}]{Garrod_ea07}
{Garrod}, R.~T., {Wakelam}, V., \& {Herbst}, E. 2007, \aap, 467, 1103

\bibitem[{{Garrod} {et~al.}(2008){Garrod}, {Weaver}, \& {Herbst}}]{Garrod_ea08}
{Garrod}, R.~T., {Weaver}, S.~L.~W., \& {Herbst}, E. 2008, \apj, 682, 283

\bibitem[{Gerlich \& Horning(1992)}]{GerlichHorning92}
Gerlich, D. \& Horning, S. 1992, Chemical Reviews, 92, 1509

\bibitem[{{Hasegawa} \& {Herbst}(1993)}]{HasegawaHerbst93}
{Hasegawa}, T.~I. \& {Herbst}, E. 1993, \mnras, 261, 83

\bibitem[{{Hasegawa} {et~al.}(1992){Hasegawa}, {Herbst}, \&
  {Leung}}]{Hasegawa_ea92}
{Hasegawa}, T.~I., {Herbst}, E., \& {Leung}, C.~M. 1992, \apjs, 82, 167

\bibitem[{{Herbst}(1980)}]{Herbst80}
{Herbst}, E. 1980, \apj, 241, 197

\bibitem[{{Herbst}(1985)}]{Herbst85}
---. 1985, \apj, 291, 226

\bibitem[{{Herbst} \& {van Dishoeck}(2009)}]{HerbstvanDishoeck09}
{Herbst}, E. \& {van Dishoeck}, E.~F. 2009, \araa, 47, 427

\bibitem[{{Hirano} {et~al.}(1999){Hirano}, {Kamazaki}, {Mikami}, {Ohashi}, \&
  {Umemoto}}]{Hirano_ea99}
{Hirano}, N., {Kamazaki}, T., {Mikami}, H., {Ohashi}, N., \& {Umemoto}, T.
  1999, in Star Formation 1999, ed. T.~{Nakamoto} (Nobeyama Radio Observatory),
  181--182

\bibitem[{{Horn} {et~al.}(2004){Horn}, {M{\o}llendal}, {Sekiguchi}, {Uggerud},
  {Roberts}, {Herbst}, {Viggiano}, \& {Fridgen}}]{Horn_ea04}
{Horn}, A., {M{\o}llendal}, H., {Sekiguchi}, O., {Uggerud}, E., {Roberts}, H.,
  {Herbst}, E., {Viggiano}, A.~A., \& {Fridgen}, T.~D. 2004, \apj, 611, 605

\bibitem[{Karpas \& Meot-Ner(1989)}]{KarpasMeotNer89}
Karpas, Z. \& Meot-Ner, M. 1989, The Journal of Physical Chemistry, 93, 1859

\bibitem[{{Katz} {et~al.}(1999){Katz}, {Furman}, {Biham}, {Pirronello}, \&
  {Vidali}}]{Katz_ea99}
{Katz}, N., {Furman}, I., {Biham}, O., {Pirronello}, V., \& {Vidali}, G. 1999,
  \apj, 522, 305

\bibitem[{{Kim} \& {Kaiser}(2012)}]{KimKaiser12}
{Kim}, Y.~S. \& {Kaiser}, R.~I. 2012, \apj, 758, 37

\bibitem[{{Marcelino} {et~al.}(2005){Marcelino}, {Cernicharo}, {Roueff},
  {Gerin}, \& {Mauersberger}}]{Marcelino_ea05}
{Marcelino}, N., {Cernicharo}, J., {Roueff}, E., {Gerin}, M., \&
  {Mauersberger}, R. 2005, \apj, 620, 308

\bibitem[{{Marcelino} {et~al.}(2009){Marcelino}, {Cernicharo}, {Tercero}, \&
  {Roueff}}]{Marcelino_ea09}
{Marcelino}, N., {Cernicharo}, J., {Tercero}, B., \& {Roueff}, E. 2009, \apjl,
  690, L27

\bibitem[{{{\"O}berg} {et~al.}(2011){{\"O}berg}, {Boogert}, {Pontoppidan}, {van
  den Broek}, {van Dishoeck}, {Bottinelli}, {Blake}, \& {Evans}}]{Oeberg_ea11}
{{\"O}berg}, K.~I., {Boogert}, A.~C.~A., {Pontoppidan}, K.~M., {van den Broek},
  S., {van Dishoeck}, E.~F., {Bottinelli}, S., {Blake}, G.~A., \& {Evans}, II,
  N.~J. 2011, \apj, 740, 109

\bibitem[{{{\"O}berg} {et~al.}(2010){{\"O}berg}, {Bottinelli}, {J{\o}rgensen},
  \& {van Dishoeck}}]{Oeberg_ea10}
{{\"O}berg}, K.~I., {Bottinelli}, S., {J{\o}rgensen}, J.~K., \& {van Dishoeck},
  E.~F. 2010, \apj, 716, 825

\bibitem[{{Occhiogrosso} {et~al.}(2011){Occhiogrosso}, {Viti}, {Modica}, \&
  {Palumbo}}]{Occhiogrosso_ea11}
{Occhiogrosso}, A., {Viti}, S., {Modica}, P., \& {Palumbo}, M.~E. 2011, \mnras,
  418, 1923

\bibitem[{{Vasyunin} \& {Herbst}(2013)}]{VasyuninHerbst13}
{Vasyunin}, A.~I. \& {Herbst}, E. 2013, \apj, 762, 86

\bibitem[{{Vasyunin} {et~al.}(2008){Vasyunin}, {Semenov}, {Henning}, {Wakelam},
  {Herbst}, \& {Sobolev}}]{Vasyunin_ea08}
{Vasyunin}, A.~I., {Semenov}, D., {Henning}, T., {Wakelam}, V., {Herbst}, E.,
  \& {Sobolev}, A.~M. 2008, \apj, 672, 629

\bibitem[{{Vasyunin} {et~al.}(2004){Vasyunin}, {Sobolev}, {Wiebe}, \&
  {Semenov}}]{Vasyunin_ea04}
{Vasyunin}, A.~I., {Sobolev}, A.~M., {Wiebe}, D.~S., \& {Semenov}, D.~A. 2004,
  Astronomy Letters, 30, 566

\bibitem[{{Wakelam} \& {Herbst}(2008)}]{WakelamHerbst08}
{Wakelam}, V. \& {Herbst}, E. 2008, \apj, 680, 371

\bibitem[{{Wakelam} {et~al.}(2010){Wakelam}, {Herbst}, {Le Bourlot}, {Hersant},
  {Selsis}, \& {Guilloteau}}]{Wakelam_ea10}
{Wakelam}, V., {Herbst}, E., {Le Bourlot}, J., {Hersant}, F., {Selsis}, F., \&
  {Guilloteau}, S. 2010, \aap, 517, A21

\bibitem[{{Watanabe} \& {Kouchi}(2002)}]{WatanabeKouchi02}
{Watanabe}, N. \& {Kouchi}, A. 2002, \apjl, 571, L173

\bibitem[{{Woon} \& {Herbst}(2009)}]{WoonHerbst09}
{Woon}, D.~E. \& {Herbst}, E. 2009, \apjs, 185, 273

\end{thebibliography}

\clearpage

\begin{landscape}

\begin{table}
  \caption{New and Other Important Gas-phase Chemical Reactions}
  \label{tbl:newr}
  \begin{tabular}{lccccc}\hline\hline
  New Reaction & $\alpha$ & $\beta$ & $\gamma$ & Temperature range (K) & Reference \\
  \hline
  CH$_{3}$O            + H              $\rightarrow$ CH$_{3}$                + OH                 & 1.6(-10)  & 0.0  & 0.0 & 300---2000 & NIST \\
  CH$_{3}$O            + O              $\rightarrow$ H$_{2}$CO               + OH                 & 1.0(-10)  & 0.0  & 0.0 & 300---2500 & NIST \\
  CH$_{3}$O            + CH$_{3}$       $\rightarrow$ H$_{2}$CO               + CH$_{4}$           & 4.0(-11)  & 0.0  & 0.0 & 300---2500 & NIST \\
  CH$_{3}$O            + CH$_{3}$       $\rightarrow$ CH$_{3}$OCH$_{3}$       + h$\nu$             & 1.0(-15)  & -3.0 & 0.0 & 10---300   & This work \\
  \hline
    Reaction & $\alpha$ & $\beta$ & $\gamma$ & Temperature range (K) & Reference\\
    \hline
  CH$_{3}^{+}$ + CO   $\rightarrow$ C$_{2}$H$_{3}$O$^{+}$   + h$\nu$                               & 1.2(-13)  & -1.3 & 0.0 & 10---100   & \cite{Herbst85}  \\
  & & & & & \cite{GerlichHorning92} \\
  O                    + C$_{2}$H$_{3}$ $\rightarrow$ CH$_{2}$CO              + H                  & 1.6(-10)  & 0.0  & 0.0 & 10---2500  & NIST \\
  O                    + C$_{2}$H$_{5}$ $\rightarrow$ CH$_{3}$CHO             + H                  & 1.33(-10) & 0.0  & 0.0 & 10---2500  & NIST \\
  H$_{3}^{+}$          + CH$_{3}$OH     $\rightarrow$ CH$_{3}^{+}$            + H$_{2}$O + H$_{2}$ & 1.8(-9)   & -0.5 & 0.0 & 10---1000  & \cite{WoonHerbst09} \\
  CH$_{3}$OH$_{2}^{+}$ + CH$_{3}$OH     $\rightarrow$ CH$_{3}$OHCH$_{3}^{+}$  + H$_{2}$O           & 1.0(-10)  & -1.0 & 0.0 & 100---300  & \cite{Anicich03}  \\
  & & & & & \cite{KarpasMeotNer89} \\
  H$_{2}$COH$^{+}$     + H$_{2}$CO      $\rightarrow$ H$_{2}$COHOCH$_{2}^{+}$ + h$\nu$             & 8.1(-15)  & -3.0 & 0.0 & 10---300   & \cite{Horn_ea04} \\
  CH$_{3}^{+}$         + HCOOH          $\rightarrow$ HC(OH)OCH$_{3}^{+}$     + h$\nu$             & 1.0(-11)  & -1.5 & 0.0 & 10---300   & \cite{GarrodHerbst06} \\
  CH$_{3}$OH           + OH             $\rightarrow$ CH$_{3}$O               + H$_{2}$O           & 4.0(-11)  & 0.0  & 0.0 & 10---100   & \cite{Cernicharo_ea12} \\
  \hline
  \end{tabular}

  Notes: a(-b) stands for $a\times10^{-b}$. $\alpha$, $\beta$ and $\gamma$ are the coefficients in the modified Arrhenius expression for the reaction rate coefficient: $k=\alpha\cdot(T/300)^{\beta}\cdot exp(-\gamma/T)$. URL for NIST Chemical Kinetics Database: http://kinetics.nist.gov/kinetics/index.jsp

\end{table}

\end{landscape}

\clearpage

\begin{table}
  \caption{Observed and best-fit modeled fractional abundances of organic species detected in cold interstellar clouds}
  \label{tbl:abu}
  \begin{tabular}{lllllll}\hline\hline
                          & \multicolumn{2}{c}{M1} & \multicolumn{2}{c}{M10} & \multicolumn{2}{c}{Observations} \\
    Species & L1689b & B1-b & L1689b & B1-b & L1689b & B1-b \\
    \hline
    HCOOCH$_{3}$      & 4.9(-16) & 4.2(-15) & 3.3(-12) & 2.0(-12) & 7.4(-10)  B & 2.0(-11) C \\
    CH$_{3}$OCH$_{3}$ & 2.4(-17) & 1.8(-12) & 1.3(-10) & 3.7(-12) & 1.3(-10) B & 2.0(-11) C \\
    CH$_{3}$CHO       & 3.2(-11) & 4.7(-12) & 6.4(-11) & 3.7(-11) & 1.7(-10) B & 1.0(-11) C \\
    CH$_{2}$CO        & 4.4(-11) & 9.0(-12) & 8.3(-11) & 3.2(-11) & 2.0(-10) B& 1.3(-11) C \\
    CH$_{3}$O         & 5.3(-14) & 8.5(-11) & 8.5(-10) & 1.5(-10) & ---                         & 4.7(-12) C \\
    H$_{2}$CO         & 9.8(-10) & 2.9(-09)  & 5.4(-08)  & 4.8(-08)  & 1.3(-09) B  & 4.0(-10) M  \\
    CH$_{3}$OH        & 5.7(-12) & 2.5(-09)  & 2.3(-08)  & 3.3(-09)  & -                           & 3.1(-09) O    \\
    \hline
  \end{tabular}

  Notes:  a(-b) stands for $a\times10^{-b}$.   B refers to \citet{Bacmann_ea12}, C refers to \citet{Cernicharo_ea12}, M refers to \citet{Marcelino_ea05}, and O refers to \citet{Oeberg_ea10}.   Times of best fits: Model M1: 1.3$\times$10$^{5}$ yr with L1689b, 1.0$\times$10$^{6}$ yr with B1-b, Model M10: 5.1$\times$10$^{5}$ yr with L1689b, 2.6$\times$10$^{5}$ yr with B1-b.

\end{table}

\clearpage

\end{document}